\documentclass[referee]{aa}
\usepackage{graphicx}
\begin{document}

\title{A review of x-ray spectral evolution in Crab pulsar}

\titlerunning{X-ray spectral evolution in Crab pulsar}

\author{M. Vivekanand\thanks{vivek@ncra.tifr.res.in}}

\institute{National Center for Radio Astrophysics, TIFR, Pune University Campus,
P. O. Box 3, Ganeshkhind, Pune 411007, India.}

\date{Received (date) / Accepted (date)}

\abstract{
Pravdo et al (\cite{PAH}) claimed that the phase resolved x-ray spectrum in 
Crab pulsar (PSR B0531+21) shows a spectral hardening at the leading edge 
of the first peak of its integrated profile (IP); this was a new and 
unexpected result. This article reanalyzes their data, as well as some 
other related data, and argues that the spectrum is as likely to be 
unvarying (i.e., neither hardening nor softening).
}

\maketitle

\section{Introduction}

Pravdo et al (\cite{PAH}; henceforth PAH) studied the power law spectrum 
of Crab pulsar 
as a function of phase within its IP, in the energy range 5 KeV to 60
KeV, using $\approx$ 86\,000 periods of data obtained by the PCA detector
aboard 
the RXTE x-ray observatory. Their photon arrival times had a resolution 
of 250 $\mu$s, allowing them to obtain the spectrum in 60 on-pulse 
phase bins, each of duration $\approx$ 335 $\mu$s, out of a total of 100 
phase bins within the IP. Their main conclusions are ``(1) the spectrum 
softens (i.e., the power law index increases) starting at the leading 
edge of the first peak until the intensity maximum of the first peak; 
(2) ``the spectrum hardens in the inter peak region; and (3) the spectrum 
softens throughout the second peak''. They point out that item (1) above 
is unexpected on the basis of existing theoretical models. They also point 
out that the power law index $\alpha$ varying as the shape of first peak 
of the IP may be an important clue for understanding the details of the 
high energy emission of the Crab pulsar. Moreover, the increase in 
$\alpha$ before the leading edge of the first peak of the IP coincides 
with the position of the radio precursor emission (Smith, \cite{FGS86}; 
Lundgren et al. \cite{LCM95}; Moffet \& Hankins \cite{MH1996}). All this 
adds to the significance of item (1) above.

This paper reanalyzes their data, as well as some other RXTE data that 
have better time resolution, and sufficient energy resolution within 
similar energy range. This paper argues that the spectral hardening is a 
weak effect in terms of the overall spectral variation across the IP; the 
spectrum is as likely to be unvarying in the leading edge of the first 
peak of the IP.

For each data file, obtained in the EVENT mode (XTE\_SE), the Good Time 
Intervals (GTI) were obtained by using the MAKETIME tool on the corresponding 
XTE filter file; the selection criterion were (a) pointing offset less than 
0.02$\degr$, (b) elevation greater than 10$\degr$, (c) all five PCUs to be 
switched on, and time since passage of RXTE satellite through the South 
Atlantic Anomaly (TIME\_SINCE\_SAA) to be greater than 30$\degr$ or less than 
0$\degr$. Next the MGTIME tool was used to merge these GTIs with those in the 
second extension of each data file, with the AND option, to produce a final 
GTI file corresponding to each data file. These data/GTI file pairs were then
input to the FASEBIN tool, along with the orbit file of the day, to obtain 
the phase resolved spectrum of Crab pulsar in 100 phase bins within the
period. In the output of FASEBIN, the Crab nebula background was subtracted 
using an off-pulse phase range of 0.1, using the FBSUB tool.

%%%%%%%%%%%%%%%%%%%%%%%%%%%%%%%%%%%%%%%%%%%%%%%%%%%%%%%%%%%%%%%%%%%%%%%%%%%%%%%%%%%%%%%
%		TABLE 1
%%%%%%%%%%%%%%%%%%%%%%%%%%%%%%%%%%%%%%%%%%%%%%%%%%%%%%%%%%%%%%%%%%%%%%%%%%%%%%%%%%%%%%%
\begin{table}[h]
\begin{tabular}[t]{cccccccc}
\hline
\hline
DATE & ID & FILE & EXP & CHN & $\tau$ & ROFF & SHIFT \\
\hline
\hline
1996 Apr 18 & 10202 & FS37\_451ee10-451f318 & 193  & 10 & 30  & 0.02105 & -0.0007 \\
            &       & FS37\_451f6b0-451fc00 & 183  & 10 & 30  & 0.02105 & -0.0004 \\
1996 Apr 20 &       & FS37\_4536db0-4537a88 & 497  & 10 & 30  & 0.02105 & -0.0005 \\
1996 May 2  & 10204 & FS37\_463b300-463bf18 & 611  & 97 & 250 & 0.02110 & 0.0 \\
            &       & FS37\_463ca70-463d5a6 & 2014 & 97 & 250 & 0.02110 & 0.0 \\
            &       & FS37\_463e1e0-463eba8 & 464  & 97 & 250 & 0.02110 & 0.0 \\
1996 Aug 23 & 10203 & FS3f\_4f83ae0-4f84862 & 3310 & 10 & 4   & 0.02090 & -0.0003 \\
            &       & FS3f\_4f851d0-4f85ee1 & 3214 & 10 & 4   & 0.02090 & -0.0003 \\
            &       & FS3f\_4f86850-4f87561 & 3214 & 10 & 4   & 0.02091 & -0.0004 \\
            &       & FS3f\_4f87ed0-4f88be1 & 3214 & 10 & 4   & 0.02093 & -0.0004 \\
\hline
\end{tabular}
\caption{
	Columns 1,2 and 3 
	contain the date of observation, the ObsID of RXTE, and the name of the 
	event mode data file. Column 4 contains the total duration of observation
	in seconds; this is obtained by summing the time intervals in the GTI file.
	In principle this is also supposed to be the time (divided by 100) found in 
	the EXPOSURE column of the phase resolved spectrum; for unknown reasons 
	this is not true for the ObsID 10204 data, but is true for the rest of the
	data. Column 5 contains the number of energy channels available within the 
	energy range analyzed; this is 5.2 keV to 59.1 keV for ObsIDs 10202 and 
	10204, and 18.2 to 54.9 keV for ObsID 10203 (in Vivekanand \cite{MV2001a} 
	and \cite{MV2001b}, which analyze data of ObsID 10203, the above energy 
	range was wrongly stated to be 13.3 Kev to 58.4 KeV). Column 6 contains 
	the time resolution of the data in micro seconds ($\mu$s). Column 7 contains 
	the offset time in seconds to be given to the FASEBIN tool. The last column 
	contains the relative shift of the IPs of each data file, with respect to 
	the IP of the data file FS37\_463ca70-463d5a6, in units of pulsar phase; for 
	comparison, the width of each phase bin is 0.01.
	}
\end{table}
%%%%%%%%%%%%%%%%%%%%%%%%%%%%%%%%%%%%%%%%%%%%%%%%%%%%%%%%%%%%%%%%%%%%%%%%%%%%%%%%%%%%%%%

Table 1 lists some information about the data analyzed in this paper. The data analyzed 
by PAH is FS37\_463ca70-463d5a6, observed on 1996 May 2. They ignored the other two 
files observed on the same day, for ``dead-time and rate considerations'', although 
they state in their appendix that ``these 8 s data are usable for spectral analysis in 
principle, since the dead-time  is spectrally independent''. The rest of the data was 
chosen from the RXTE public data archive using the criterion: (1) the number of usable 
energy channels within 5 KeV to 60 Kev must be at least 10, for proper fitting of the 
power law spectrum, and (2) the time resolution should be much smaller than the width of 
each phase bin (which is $\approx$ 335 $\mu$s). PAH analyze data whose accuracy is 
comparable to the width of their phase bins; ideally the former should be much smaller 
than the latter. A fourth data file in ObsID 10202, and data in ObsID 40090, had the best 
time resolution (2 $\mu$s), but $\le$ 8 usable energy channels.

The output of the FASEBIN tool is the required phase resolved spectrum. This tool uses 
two parameters that can potentially affect further analysis -- the Crab pulsar period 
$P$ for the epoch of observation, and the time offset ROFF (column 7 of table 1) with 
respect to the radio ephemeris. The former is automatically obtained from the pulsar 
catalog, while the latter has to be inserted by hand. Now, the $P$ of epoch for Crab 
pulsar can be different from the catalog determined value owing to its glitching and 
timing activity; and an error of 3.5 nano second in the period used for folding can 
cause a total drift of 1 phase bin in one of the longer data files in table 1. Next, 
the data are obtained not in one contiguous data file, but in several files, each 
observed at different epochs. Therefore the correct time offset ROFF for each file is 
crucial to align in phase the data in those files; otherwise the $i^{\mathrm{th}}$ 
(say) phase bin in each file might correspond to different true phase bins within the 
pulsar period. 

To avoid these problems, two things were done. First, the ASCII version of the photon 
counts were obtained from the FITS version of each phase resolved spectrum. An IP was 
formed from this data, combining all energy channels. This was cross correlated with 
a standard IP (that of data file 10204:FS37\_463ca70-463d5a6, which itself was first 
suitably shifted in phase for centering it). Any significant phase shift detected was 
added to ROFF, and the FASEBIN tool was run once again. In this iterative manner it 
was ensured that the residual phase shift (column 8 of table 1) is much less than the 
width of a phase bin (0.01). 

Next, arrival times of photons in each data file above were converted to the solar 
system barycenter system (TDB) as described in Vivekanand (\cite{MV2001a}, \cite{MV2001b}), 
using the new FAXBARY tool (improved version in HEASOFT 5.1). Then the DECODEEVT tool 
was used to obtained the TDB times of each photon event, and its energy channel. From
the ASCII version of this data, IPs were formed of the first and second halves of 
each data file above. These two were then cross correlated. Their relative phase shift 
was sufficiently small to ensure that the period used was sufficiently accurate for 
the current purpose (see Vivekanand \cite{MV2001a}).

Penultimately, the response matrix for each data file was obtained using the PCARSP 
tool. Since the phase resolved spectrum output by FASEBIN is incompatible with PCARSP,
a non-phase resolved spectrum was obtained for each data file using the tools SEFILTER
and SEEXTRCT, using the corresponding GTI file.

Finally, the background subtracted phase resolved spectra were analyzed using the XSPEC 
tool. The spectrum was modeled as the power law $dN/dE = \beta E^{-\alpha}$ (photons 
per keV), where $\beta$ is the normalization constant and $\alpha$ is the power law 
index. Only energy channels lying between 2 keV and 60 keV were considered, since this 
is the RXTE/PCA energy calibrated range. The channels to ignore were obtained by 
looking at the corresponding response matrices using FDUMP.

\section{Reanalysis of data of Pravdo et al (1997)}

Figure 1 shows the result of fitting, to the data of PAH, the power law model 
spectrum to 66 of the 100 phase bins, that are centered on the Crab pulsar x-ray 
emission. The top panel of fig.~1 shows the background subtracted Crab pulsar 
emission, which is in agreement with their profile in their fig.~1. 

The middle panel of fig.~1 shows the variation of $\beta$ with pulse phase. This
is in qualitative agreement with the $\beta$ values tabulated by PAH; but these are 
systematically higher by the factor $\approx$ 1.5. This author does not understand 
this factor, and neither do the people at the XTE Help Desk, who are addressing 
the problem. The current suspicion is that the total duration of observation 
(contained in the EXPOSURE column of the phase resolved spectrum) is erroneous for 
data of this ObsID; For this file the EXPOSURE is indicated to be 2870 secs, while
it is actually 2014 sec (table 1). Please note that this problem exists for this 
ObsID only. However, internal consistency has been verified by integrating, at 
several phases within the IP, the derived power law within a given range of 
energies, and comparing this with the corresponding result of the FLUX command 
in XSPEC.

The last panel of fig.~1 shows the variation of $\alpha$ with pulse phase. Once 
again these agree qualitatively very well with the $\alpha$ values of PAH. The 
overall behavior is very similar, although there are differences in detail.

Figure 3 of PAH shows spectra at two phases within the IP, with the model spectrum
superimposed upon the data. This figure was also reproduced in the current analysis,
and there is very good qualitative agreement between the two plots.

The normalized $\chi^2_d$ (i.e., $\chi^2$ per degree of freedom) is distributed 
similarly in this work and that of PAH. In the current work $\chi^2_d$  lies 
between 0.7 and 1.4 in the 66 phase bins in which spectral estimation was possible. 
In PAH the range is exactly the same in the 60 phase bins analyzed. In the current 
work 25 of the 66 $\chi^2_d$ have values $\ge$ 1.1, while the corresponding numbers 
in PAH are 30 out of 60. The number of $\chi^2_d$ $\ge$ 1.2 in the two works is the 
same, viz., 14.

It is therefore concluded that this analysis broadly reproduces the results of PAH.

\begin{figure}
\resizebox{\hsize}{!}{\includegraphics{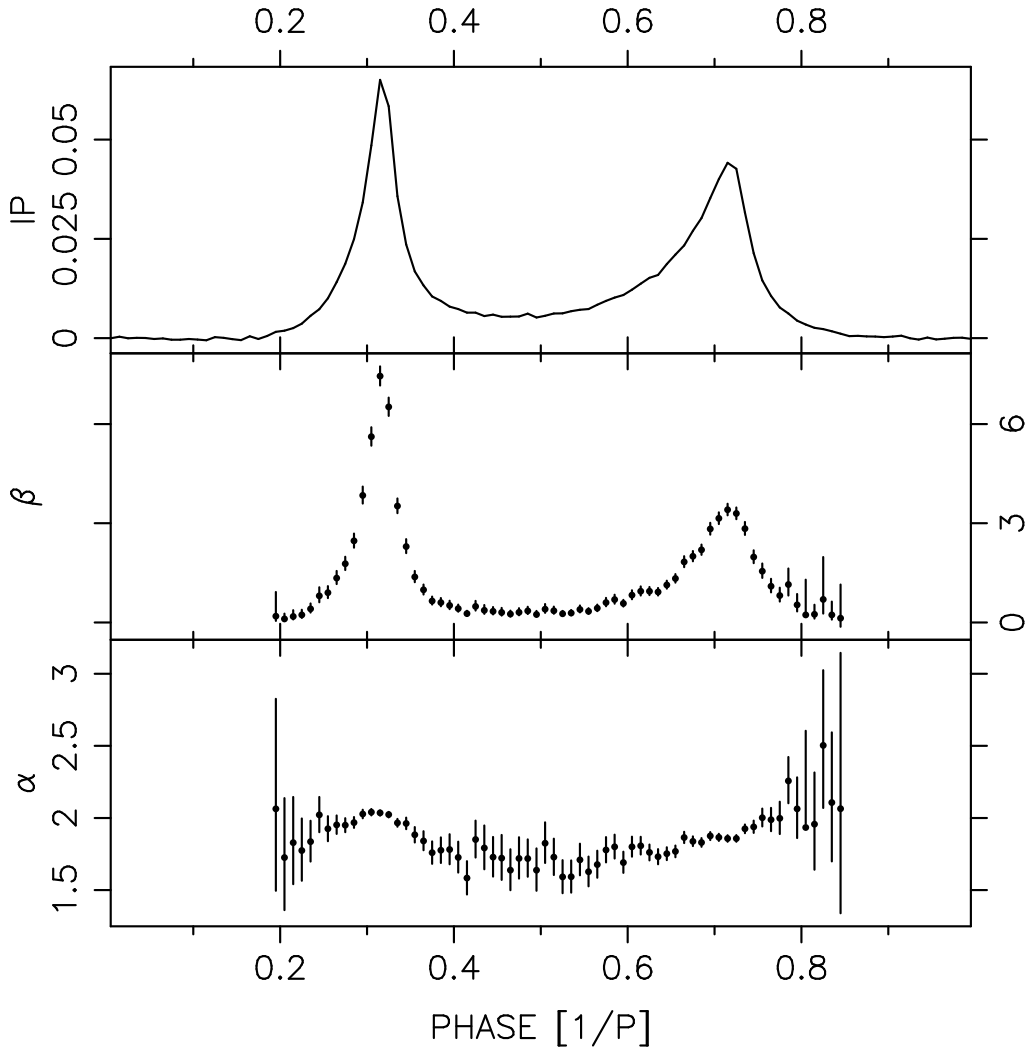}}
\caption{
	Reanalysis of the data file FS37\_463ca70-463d5a6, ObsID 10204, observed on 
	1996 May 2 (Pravdo et al \cite{PAH}). {\bf Top panel}: IP of Crab pulsar. 
	The Crab nebula emission has been subtracted by using FBSUB and the two 
	phase ranges 0.01-0.05, and 0.96-1.00.; the area under the curve has been 
	normalized to 1.0. {\bf Middle panel}: The normalization constant $\beta$. 
	{\bf Bottom panel}: The power law index $\alpha$. In the lower two panels, 
	the dots represent the mean values, while the vertical bars represent the 
	limits at the 90\% confidence level, as derived by XSPEC. The range of
	energy used is 5.2 keV to 59.1 keV. The fit at phase 0.81 is dubious.
        }
\label{fig1}
\end{figure}

Some minor differences exist between this analysis and that of PAH.  In this 
analysis, a range of 0.1 in phase is used for nebular background subtraction 
(0.01-0.05, and 0.96-1.00), while the latter use a range of 0.3. Next, this work 
quotes the 90\% confidence limits on $\alpha$ and $\beta$, while the latter quote 
the one standard deviation limits. Next, the boundaries of the phase bins will 
not match exactly in the two cases, since the centering of the Crab emission 
within the IP is done differently (in fact, arbitrarily) in the two cases. In 
this analysis the exact energy range chosen for analysis is 5.2 keV to 59.1 keV 
(ignored energy channels 98 on wards); the corresponding numbers have not been 
mentioned by PAH. For example, if they included the next energy channel, their 
energy range would go up to 60.4 keV.

Other differences can also exist. For example, PAH do not mention how they obtained 
their GTIs. If they did not choose the selection based on TIME\_SINCE\_SAA, which is
apparently often done to maximize scarce data, it will make a difference, although 
very small, to the amount of data retained for analysis.

Given these actual and possible differences, one does not expect the current results
to match identically with those of PAH. So Fig.~1 will be taken as a qualitative 
confirmation of their broad results. One can see from this figure why they thought 
that there was spectral hardening at the leading edge of the first peak of the IP 
(phase range 0.19-0.32). If one ignores the data at phases 0.19 and 0.24, one can 
discern a clear hardening of the spectrum (i.e., decrease in $\alpha$) at earlier 
phases with respect to the first peak (which falls at phase 0.32 in fig.~1). PAH 
do not do any quantitative analysis in this regard, presumably because of the 
non-Gaussian nature of the distribution of errors.

However, one can notice that most of the visual effect of spectral hardening comes 
from the the four data at phases 0.20-0.23; unfortunately these have much higher 
error bars. The four data from phases 0.25-0.28 do not show such dramatic spectral 
hardening; they could even be consistent with a flat distribution.

This section will be concluded with the inspiration that in fig.~1 there is 
certainly a {\bf visual} case for spectral hardening at the leading edge of the 
first peak of the x-ray IP of Crab pulsar, as noticed by PAH in the same data, not 
withstanding the minor differences between the two works.

In the next section it will be argued that such an inspiration could be misleading.

\section{Analysis of data that Pravdo et al (1997) ignored}

Figure 2 displays the IP and the $\alpha$ for the other two data files of ObsID 10204,
which were ignored by PAH. The method of analysis is identical to that described
earlier. The IP is almost exactly the same as in fig.~1, and broadly $\alpha$ varies 
with phase as in fig.~1. The normalized $\chi^2_d$ lies between 0.8 and 1.3 in the 66 
phase bins in which spectral estimation was possible; 29 of these $\chi^2_d$ have 
values $\ge$ 1.1, while 13 of them have values $\ge$ 1.2. Therefore the $\chi^2_d$ 
in fig.~2 appears to be distributed as in fig.~1, as well as in PAH. The total 
duration of observation in fig.~2 is more than half of that of fig.~1, so one can 
certainly compare the $\alpha$ variation in the two figures.

In fig.~2 there is hardly any spectral hardening in the phase range 0.19-0.32. Now 
one has to ignore the data at three phases (0.19, 0.20 and 0.22) to discern the 
spectral hardening seen in fig.~1. However, the $\chi^2_d$ for these three data 
are 1.25, 1.10 and 1.15; so at least the latter two data are certainly well 
determined. What is most significant is the almost flat $\alpha$ distribution in 
the phase range 0.26-0.32. This contrasts with the bump seen in the bottom panel 
of fig.~1 in the same phase range. In fact, if one were allowed to disregard the 
single $\alpha$ at phase 0.29 in fig.~2, then one can probably discern a trough, 
instead of a bump, in the phase range 0.26-0.32! If one ignores none of the 
$\alpha$ below phase 0.32 in fig.~2, then there appears to be more of a case for 
a {\bf flat} $\alpha$ distribution, rather than it decreasing with decreasing 
phase.

\begin{figure}
\resizebox{\hsize}{!}{\includegraphics{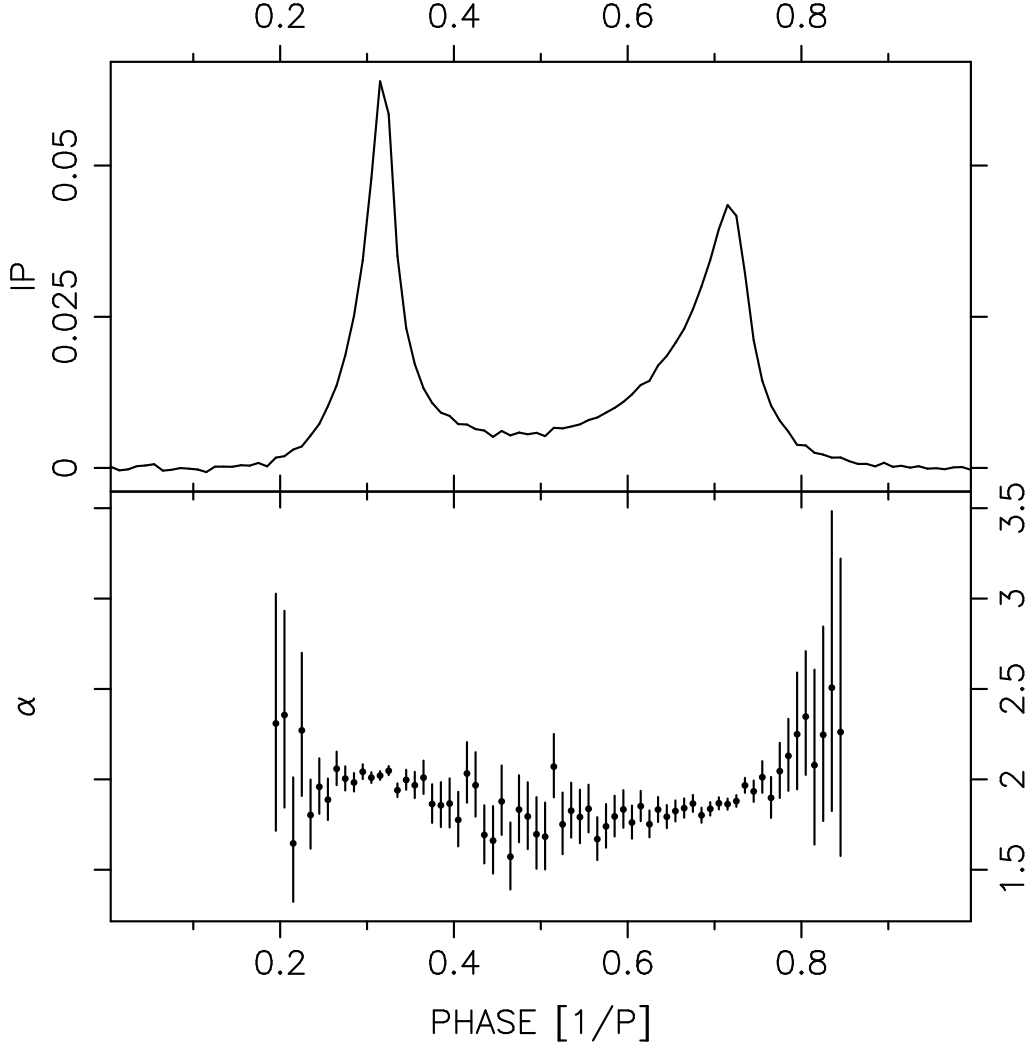}}
\caption{
        Analysis of data files FS37\_463b300-463bf18 and FS37\_463e1e0-463eba8, ObsID 
	10204, observed on 1996 May 2 (this data was ignored by Pravdo et al \cite{PAH}). 
	{\bf Top panel}: IP of Crab pulsar. The Crab nebula emission has been subtracted 
	by using FBSUB and the two phase ranges 0.01-0.05, and 0.96-1.00.; the area under 
	the curve has been normalized to 1.0. {\bf Bottom panel}: The power law index 
	$\alpha$. The dots represent the mean values, while the vertical bars represent 
	the limits at the 90\% confidence level, as derived by XSPEC. The range of energy 
	used is 5.2 keV to 59.1 keV.
        }
\label{fig2}
\end{figure}

It is unfortunate that one is unable to do a quantitative analysis like the standard
$\chi^2$, due to the non-Gaussian nature of the errors. However, at the level of the
{\bf visual} analysis done by PAH, this section concludes that the data they ignored
does not appear to show the spectral hardening that they claim. The data of fig.~2
have been analyzed in the same manner as that in fig.~1. So whatever may be the
differences between fig.~1 here and the results of PAH, one expects consistency 
between figs. 1 and 2 of this paper. It is predicted that if PAH were to analyze the 
data they have ignored, in the same manner as in PAH, they will reproduce a figure 
similar to fig.~2 here.

\section{Analysis of other data}

\subsection{ObsID 10202}

Figure 3 shows the results for three files of ObsID 10202. The reason they were
not combined is the following. First, they belong to different sets of
observations (10202-02-01-00 and 10202-02-02-00); from their response files one
notices small differences in the labeling of their energy channels. Second,
there was some technical problem in obtaining the response matrix for the data
file FS37\_451f6b0-451fc00 (by the method mentioned in section 1). So it was
obtained by not using the attitude file. It was verified that the numerical 
values in each column of this response matrix are very similar to those in the
response matrix for the other file (FS37\_451ee10-451f318) in this observation 
set; the maximum difference was 0.3\%.

In the bottom panel of fig.~3, $\alpha$ appears to behave as in fig.~2, rather
than as in fig.~1. The bump in the phase range 0.26-0.32 is less prominent 
than in fig.~1, and $\alpha$ at the smallest phases appears to be higher
rather than lower. Even if one ignores the data at phase 0.20, which has a
high $\chi^2_d$ of 1.9, the spectral hardening claimed by PAH is weak, at 
best. The $\chi^2_d$ vales in this panel range from 0.2 to 3.0; the numbers 
lying $\ge$ 1.1 and $\ge$ 1.2 are 27 and 21, respectively, out of the 61 
phase bins analyzed. This is similar to the situation in figs. 1 and 2.

The spectral hardening is weaker, and probably non-existent, in the 
middle panel of fig.~3. The $\chi^2_d$ distribution in this panel is similar
to other distributions.

\begin{figure}
\resizebox{\hsize}{!}{\includegraphics{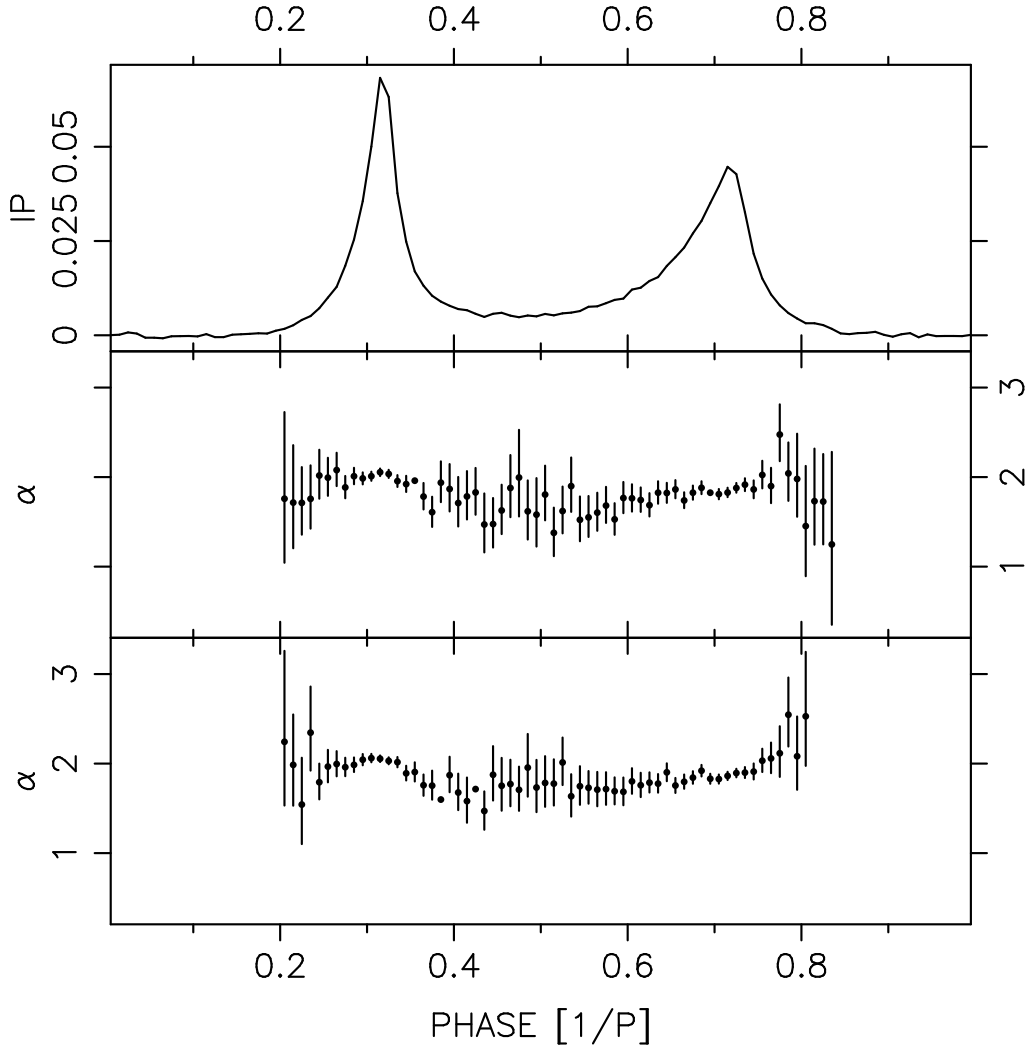}}
\caption{
        Analysis of data of ObsID 10202. {\bf Top panel}: Same as the top panel in 
	fig.~1 but for data files FS37\_451ee10-451f318 and FS37\_451f6b0-451fc00. 
	{\bf Middle panel}: Same as the bottom panel in fig.~2, but for data files 
	FS37\_451ee10-451f318 and FS37\_451f6b0-451fc00. {\bf Bottom panel}: Same 
	as the bottom panel in fig.~2, but for data file FS37\_4536db0-4537a88. The 
	range of energy used is 5.2 keV to 59.1 keV.
        }
\label{fig3}
\end{figure}

\subsection{ObsID 10203}

Figure 4 shows the results for four data files of ObsID 10203; their 
combined observing time is much larger than that of any other set in table 1. 
The individual $\chi^2_d$ of the first eight data from phases 0.19 to 0.26 
does not exceed 1.16; so these $\alpha$ values are very reliable. For the 65 
phases analyzed in this figure, the $\chi^2_d$ range from 0.6 to 2.1; 24 of 
these have values $\ge$ 1.1 while 14 of these values $\ge$ 1.2. If one could 
ignore the $\alpha$ at phase 0.21, one might argue that a strong case exists 
for spectral hardening; but if one could ignore the first two $\alpha$ also, 
then the case becomes much weaker. If one could ignore the first five data 
(phases 0.19-0.23) then the case becomes really weak. Even if one ignores 
the first eight data, it is not clear how strong the case is for spectral
hardening, if one takes into account the error bars on the four data at
phases 0.27-0.30.

\begin{figure}
\resizebox{\hsize}{!}{\includegraphics{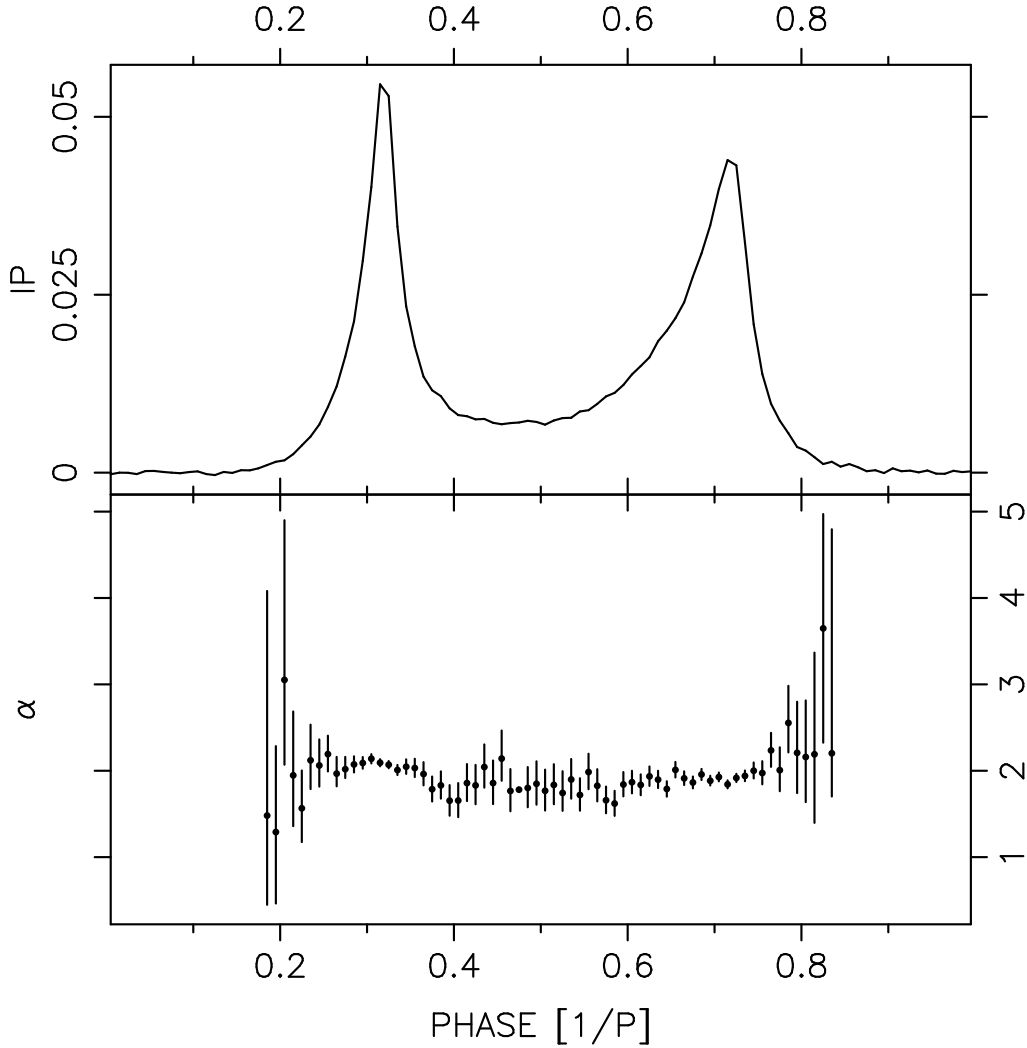}}
\caption{
        Analysis of data of ObsID 10203. {\bf Top panel}: Same as the top panel in
        fig.~2 but for the four data files of this ObsID. {\bf Bottom panel}: Same
        as the bottom panel in fig.~2, but for the four data files of this ObsID.
        The range of energy used is 18.2 keV to 54.9 keV.
        }
\label{fig4}
\end{figure}

\section{Discussion}

This paper therefore concludes that the spectral hardening noticed by PAH, at 
the leading edge of the first peak of Crab pulsar's x-ray IP, is not a strong 
result, although PAH were probably not unjustified in arriving at such a 
conclusion from their own analysis. Other data, particularly that which was 
part of the data set of PAH but which they ignored, indicates that a constant 
$\alpha$ is as likely a solution in contrast to a decreasing $\alpha$.

Apparently the above spectral hardening has also been noticed by Massaro et 
al (\cite{MCLT2000}) in the energy range 1.6 KeV to 300 KeV, using the BeppoSAX 
observatory. They claim in their discussion that regarding the first peak of 
the IP, they have `confirmed ...  that both the leading and trailing edges 
have harder spectra than the central bins'. In particular they note that in
their figs. 3c and 3d, ``the central bins'' of the first peak ``have a softer
spectrum than the wings''; these figures are similar to the bottom panel of 
fig.~2 here, but for energy ranges 10 keV to 34 keV, and 15 keV to 300 keV,
respectively. However, a closer look at these two figures of Massaro et al 
(\cite{MCLT2000}) indicates that this result may be be non-existent in their 
fig.~3d; $\alpha$ appears to be more or less constant at phases below that 
of the first peak of the IP. Their fig.~3c certainly contains what might be 
interpreted as the spectral hardening noticed by PAH. If one were to take 
both these results seriously, one might be forced to the interpretation that 
the spectral hardening, if at all it exists in the BeppoSAX data, is likely 
to be a phenomenon confined to a narrow energy range (10 keV to 34 keV). 

Figure 3b of Massaro et al (\cite{MCLT2000}), energy range 1.6 keV to 10 keV, 
shows that $\alpha$ may be constant at those phases where it decreases the 
most in their fig.~3c; and it shows a small bump where it is more or less 
constant in their fig.~3c. Finally their fig.~4a (15 KeV to 80 KeV range) 
appears to show a nearly unvarying $\alpha$ at the leading edge of the first 
IP peak.  It therefore appears that the results of Massaro et al 
(\cite{MCLT2000}) are also indicating the kind of confusion that has been 
presented in this paper, regarding the spectral hardening claimed by PAH.

At this stage one can certainly not rule out the spectral hardening claimed
by PAH, particularly since another independent instrument claims to have
noticed the same. However, one should also keep in mind the divergence of
results in this regard, and also not rule out the possibility that $\alpha$
may be more or less constant in the leading edge of the first peak of the
IP. This is particularly important in view of (a) the lack of quantitative
analysis in this regard, and (b) a recent theoretical study prefers a
monotonically {\bf increasing} $\alpha$ at smaller phases within the IP,
and neither a constant nor a decreasing $\alpha$ (Zhang and Cheng
\cite{ZC2002}). To accommodate the spectral hardening of PAH, they need to
invoke emission other than synchrotron radiation from secondary pairs.

Pravdo et al (\cite{PAH}) claim that phase resolved x-ray spectrum in Crab
pulsar ``softens throughout the second peak'' of the IP (their fig.~1).
However fig.~\ref{fig1} of this paper shows that this is also a weak effect, 
as do fig.~\ref{fig4} and the bottom panel of fig.~\ref{fig3}; in these 
figures the spectrum could be unvarying in the trailing edge of the second 
peak also. The above claim of PAH appears to be more evident in 
fig.~\ref{fig2} of this paper. However, in the second panel of 
fig.~\ref{fig3}, the spectrum appears to {\bf harden} in the trailing edge
of the second peak, opposite to the claim of PAH. Keeping in mind the lack
of quantitative analysis, and the confusing visual results from different
data sets, one can not rule out a constant $\alpha$ in the trailing edge
of the second peak of the x-ray IP of Crab pulsar.

\begin{acknowledgements}

I thank Teresa Mineo and the anonymous referee for useful suggestions. This 
research has made use of (a) High Energy Astrophysics Science Archive Research 
Center's (HEASARC) facilities such as their public data archive, and their 
FTOOLS software, and (b) NASA's Astrophysics Data System (ADS) Bibliographic 
Services.

\end{acknowledgements}


\begin{thebibliography}{}

\bibitem[1995]{LCM95}
Lundgren, S. C., Cordes, J. M., Ulmer, M., Matz, S. M., Lomatch, S., Foster, R. S.,
Hankins, T., 1995, ApJ, 453, 433

\bibitem[2000]{MCLT2000}
Massaro, E., Cusumano, G., Litterio, M., Mineo, T. 2000, A\&A, 361, 695

\bibitem[1996]{MH1996}
Moffet, D. A.,  Hankins, T. H., 1996, ApJ, 468, 779

\bibitem[1997]{PAH}
Pravdo, S. H., Angelini, L., Harding, A. K., 1997, ApJ, 491, 808 (PAH)

\bibitem[1986]{FGS86}
Smith, F. G., 1986, MNRAS, 219, 729

\bibitem[2001a]{MV2001a}
Vivekanand, M, 2001, A\&A, 373, 236

\bibitem[2001b]{MV2001b}
Vivekanand, M, 2001, A\&A, 376, 580

\bibitem[2002]{ZC2002}
Zhang, L., Cheng, K. S., 2002, ApJ, 569, 872

\end{thebibliography}
\end{document}